# Beyond Residence: A Mobility-based Approach for Improved Evaluation of Human Exposure to Environmental Hazards

Zhewei Liu[1]*, Chenyue Liu[1], Ali Mostafavi[1]

[1] UrbanResilience.AI Lab, Zachry Department of Civil and Environmental Engineering, Texas A&M University, College Station, TX, 77843

**Abstract:**
Evaluating human exposure to environmental hazards is crucial for identifying susceptible communities and devising targeted health policies. Standard environmental hazard exposure assessment methods have been primarily based on place of residence, an approach which neglect individuals' hazard exposures due to the daily life activities and mobility outside home neighborhood. This limitation has led to underestimation of peoples' overall hazard exposure risk. To address this limitation, this study proposes a novel mobility-based index for hazard exposure evaluation. Using large-scale and fine-grained human mobility data, we quantify the extent of population dwell time in high-environmental-hazard places in 239 U.S. counties for three major environmental hazards: air pollution, heat, and toxic sites. Subsequently we explore the extent to which human mobility extends the reach of environmental hazards and also lead to the emergence of latent exposure for populations living outside high hazard areas with relatively considerable dwell time in high hazard areas. The findings help quantify environmental hazard exposure more reliably, considering the role of human mobility and activities. Notably, the neglect of mobility can lead to a 10% underestimation of hazard exposure risks, particularly in areas with direct exposure. The interplay of spatial clustering in high-hazard regions and human movement trends creates "environmental hazard traps" intensifying exposure. A notable disparity was observed in exposure levels within communities, confirming dire environmental injustices. Poor and ethnic minority residents disproportionately face multiple types of environmental hazards, aggravating potential health impacts. This data-driven evidence supports the severity of these injustices. We also studied latent exposure arising from visits outside residents' home areas, revealing millions population having 5% to10% of daily activities occur in high-exposure zones. Despite living in perceived safe areas, human mobility could expose millions of residents to different hazards. These findings provide crucial insights for targeted policies to mitigate these severe environmental injustices.

## 1. Introduction

Environmental hazards, such as air pollution, toxic exposures, and heat, have become pressing concerns in the face of rapid urbanization, industrialization, and climate change. The adverse effects of hazard exposures on populations worldwide are increasingly recognized, with an estimated 23% of global deaths, or roughly 12.6 million deaths per year, attributable to environmental factors (Prüss-Üstün, Wolf et al. 2016, Guo, Shi et al. 2022). Global climate change further aggravates the detrimental health consequences of environmental hazard exposure, underscoring the importance of accurate hazard monitoring and evaluation. (Landrigan, Fuller et al. 2018, Romanello, McGushin et al. 2021).

Evaluating and monitoring environmental hazards and their impact on human populations has been the subject of numerous studies in recent years. Traditional methods for assessing hazards and populations exposure have been primarily focused on the place of residence, using empirical models, environmental data from sensors, and spatial analysis techniques to estimate hazard concentrations in different areas of a city (Reid, O'neill et al. 2009, Yin, Grundstein et al. 2021, Huang and Mostafavi 2023). Researchers have proposed various indicators to quantify the extent of environmental hazards exposures and risks of communities. For instance, researchers have estimated PM2.5 exposure by assigning emissions from industrial facilities to nearby census block groups and then incorporating this pollutant into calculations of the burden on racial groups and poverty status according to their residence locations (O'Neill, Zanobetti et al. 2005, Johnson and Wilson 2009). Likewise, heat vulnerability indices (HVI) have been developed to assess heat risk at specific locations based on the populations' social-economic factors, land cover type, and green space (Wolf and McGregor 2013). These indices reveal populations at higher exposure locations experiencing greater mortality rates during periods of high temperatures (Reid, Mann et al. 2012, Maier, Grundstein et al. 2014).

However, the existing approaches tend to overlook the effects of daily activity and movement patterns, which can significantly alter an individual's exposure to various hazards. The consideration of a mobility-based evaluation of hazard exposure provides a deeper understanding of the extent to which individuals and communities are exposed to different types of environmental hazards. The traditional residence-based methods for assessing hazard exposure do not adequately account for the fact that people's daily activities involve movements between different locations with varying levels of hazard exposure (Kuras, Hondula et al. 2015, Kuras, Richardson et al. 2017). For instance, daily commuting patterns can lead to increased exposure to air pollution, even for individuals who reside in areas with relatively low levels of pollution (Fan, Chien et al. 2022). Similarly, exposure to urban heat can vary significantly throughout the day due to differences in land use, building materials, and the presence of green spaces, which can impact people who frequently visit parks, retail establishments, or attend sporting events (Yin, Grundstein et al. 2021, Li, Jiang et al. 2023). Lack of consideration of exposures caused by human mobility has created blind spots and also undermines the accuracy of hazard exposure assessments.

Another important aspect of hazard exposure research is the environment justice issue, which emphasizes that certain population groups and communities may be disproportionately exposed to the adverse effects of environmental hazards (Brulle and Pellow 2006). Studies have shown that minority and low-income communities often face a higher burden of hazard exposure, such as air pollution and extreme heat, resulting in health disparities and social inequity (Clark, Millet et al. 2014). Research on air pollution exposure reveals that racial and ethnic minorities and low-income individuals are more likely to live in areas with higher concentrations of PM2.5 emissions, leading to increased health risks (Bell and Ebisu 2012). Similarly, low-income and minority communities are often more susceptible to the impacts of heatwaves due to the lack of access to green spaces, inadequate housing, and limited resources for adaptation (Harlan, Brazel et al. 2006).

Similar to the standard environmental hazard evaluations, the majority of environmental justice studies have mainly focused on residence-based analyses of hazard exposure disparities (LeBrón, Torres et al. 2019). However, with recognition of the importance of human mobility in determining the extent of hazard exposure of populations, there is an increasing need to explore environmental justice issues through this lens. The fine-scaled human mobility datasets enable tracking daily activities at individual and crowd levels (Wang, Zhang et al. 2020, Liu, Wang et al. 2022), as well as revealing the urban structures (Liu, Zhou et al. 2018, Liu, Zhou et al. 2019). By capturing the dynamic nature of daily activities and movements, mobility-based evaluations can provide a more accurate quantification of the disparities in hazard exposure experienced by vulnerable populations (Chakraborty, Maantay et al. 2011, Kwan 2012). An example is that low-income workers who commute to industrial sites for work may face higher exposure to air pollution, even if they reside in areas with relatively low pollution levels (Fan, Chien et al. 2022). In addition, individuals who work in urban areas with limited green spaces may experience increased vulnerability to heat stress during the day Integrating mobility-based evaluations into environmental justice research can deepen the understanding of the extent of latent exposure in these previously overlooked vulnerable communities, informing targeted interventions and policies to reduce disproportionate hazard exposure and its associated health and social impacts.

This study introduces a novel mobility-based index for hazard evaluation. Focusing on the U.S. coastal areas as study regions, we assess the impact of three significant natural hazards (i.e., air pollution, heat, and toxic sites) within these regions. Accordingly, this study aims to answer the following interrelated research questions:

- RQ1: To what extent do individuals' mobility increase their daily hazard exposure, in addition to the exposure at their residence?
- RQ2: How much population face latent hazard exposure due to their daily mobility, despite living in no-exposure areas? Which hazards pose the greatest latent threat to the populations?
- RQ3: How do hazards differentially impact various geographic regions and diverse demographic groups? How effective is a mobility-based approach in revealing under-identified communities and exposing emerging environmental issues?

## 2. Dataset and Methodology:

The workflow of this study is presented in Figure 1. First, human mobility datasets are overlaid with the distributions of the targeted hazards to quantify the dwell time of users with high-hazard areas and then aggregate the dwell time exposures in creating the mobility-based hazard exposure index. We calculate the mobility-based exposure index for each hazard type at the census tract level across different U.S. coastal counties. Second, the index is examined to reveal the extent to which human mobility extends environmental hazard exposures. Also, the mobility-based exposure index is analyzed in conjunction with demographic data to identify disparities in hazard exposure across the study regions. Third, we investigate the spatial distribution patterns and disparity in mobility-based exposure for various hazards, along with their corresponding correlations and inequalities among different sub-populations.

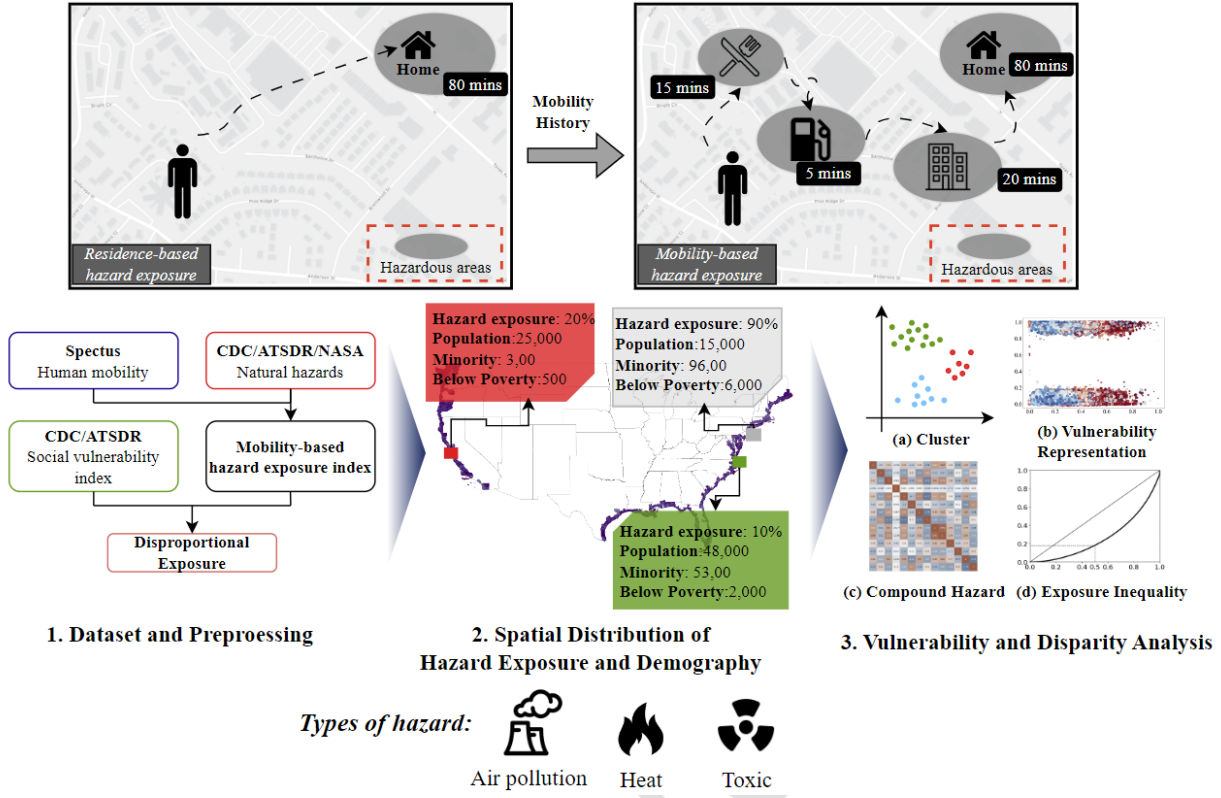

**Figure 1**. Overview of study workflow. Human mobility datasets are combined with environmental hazard datasets, to calculate mobility-based hazard exposure index based on user dwell-times in high-hazard areas. Demographic analysis is conducted to quantify the threats of hazard exposures to different sub-populations across the study regions. Third, by using clustering, correlation and inequality analysis, the vulnerability of different communities to hazard exposures and the corresponding disparity issues are revealed.

## 2.1 Datasets

We selected the coastal areas in United States as the study regions, covering 239 counties in total. The socioeconomic and hazard exposure data are collected and analyzed at the census tract level. The details related to each dataset is presented below.

*Human mobility datasets*

The human mobility datasets are used for tracking individuals' daily movement. In this study, the datasets are provided by Spectus Inc. (Spectus 2022), which includes the anonymized location of mobile phones and smartphone devices. The datasets are collected in accordance with privacy practices, ensuring the collection of anonymous and privacy-compliant location data. The dataset utilized in this study pertains to the entire month of April 2019; each record in the database includes an individual's visit history:

$$rcd_k = \{user_k, loc_k, date_k, time_k, dtime_k\} \quad (1)$$

where, $rcd_k$ is a record in our database, $user_k$ is a logged individual in the dataset, $loc_k$ is the location of a stop by $user_k$, and $date_k, time_k, dtime_k$ are respectively date, time and dwell time at the stop. The home location of each individual $user_k.home$ is inferred in accordance with previous works (Fan, Chien et al. 2022).

*Demographic datasets and Social Vulnerability Index*

Demographic data at census tract level are collected from the Social Vulnerability Index (ATSDR 2022) published by The Centers for Disease Control and Prevention's Agency for Toxic Substances and Disease Registry (CDC ATSDR). The data covers the demographic statistics such as population, average income and also the indicators for assessing a community's potential susceptibility to hazards, such as percentage of ethnic minority, poverty, lack of access to transportation, and crowded housing.

*Hazard exposure datasets*

Our study necessitates the identification of high-hazard exposure areas, for which we consider three primary environmental hazard types: air pollution, proximity to toxic sites, and extreme heat.

For air pollution, the datasets are collected from the Environmental Justice Index (ATSDR 2023), which includes the percentile rank of annual mean days exceeding the PM2.5 regulatory standard, averaged over three years, at the census tract level. A threshold of 0.5 was established for each census tract to specify high air pollution exposure areas.

For toxic exposure, the data specifies the percentile rank of the proportion of a tract's area located within a one-mile buffer of an EPA Toxic Release Inventory, again collected from the Environmental Justice Index (ATSDR 2023). A threshold of 0.5 was set for each census tract to demarcate areas with high exposure to toxic sites.

For extreme heat exposure, the data are provided by the North American Land Data Assimilation System (NASA 2022), which represents the number of extreme heat days occurring between May and September 2019. For each county, we employed quartiles to delineate five classes, subsequently classifying the top quartile, or 25%, of census tracts as high heat exposure areas.

## 2.2 Calculating mobility-based hazard exposure index

The calculated mobility-based hazard exposure index at census tract level represents an individuals' hazard exposure with consideration of both their residence location and daily dwell time in high hazard areas.

First, for a census tract $ct_i$, the parameter total dwell time $TDT_i$ is computed by summing the dwell time of each visit made by individuals who reside within the tract:

$$TDT_i = \sum_{k=1}^{n} rcd_k.dtime_k, (rcd_k.user_k.home \in ct_i) \quad (1)$$

Then, the parameter dwell time in hazards $HDT_k$ is defined as the sum of the dwell times recorded at stop points located within the high-hazard exposure areas (see Section 2.1):

$$HDT_i = \sum_{k=1}^{n} rcd_k.dtime_k, (rcd_k.user_k.home\ ct_i\ \&\ rcd_k.loc_k\ high\ exposure\ areas) \quad (2)$$

Finally, for a census tract $ct_i$, the mobility-based hazard exposure index $MEI^i$ is calculated by:

$$MEI^i = \frac{HDT_i}{TDT_i} \quad (3)$$

We respectively calculate the tract-level $MEI^i$ for all the three hazard types, and consequently calculate the mobility-based hazard exposure index ($MEI^i_{ap}$, $MEI^i_t$, $MEI^i_h$) for air pollution, toxic site exposures, and extreme heat, respectively The index captures the extent to which users residing in a census tract spend time by visiting places in high hazard areas as a proportion of total time they spent in all places.

The index range is from 0 to 1. The greater the index value, the greater the dwell in high hazard areas and the greater the exposure.

## 2.3 Classifying the regions based on mobility-based hazard exposures

To identify the patterns of various environmental hazard exposures, a density-based clustering method (DBSCAN) was applied to classify the study regions into different categories (Ester, Kriegel et al. 1996). DBSCAN is robust to data noise and can detect clusters with arbitrary shapes and sizes without assuming spherical clusters and specified number of clusters.

Given a dataset $X$ consisting of n data points, DBSCAN defines a neighborhood around each data point $x_i$ using a distance metric $d(.,.)$ and a radius $\varepsilon > 0$:

$$N_\varepsilon(x_i) = \{x_j \in X : d(.,.) \leq \varepsilon\} \quad (4)$$

where $N_\varepsilon(x_i)$ is the set of points within the $\varepsilon$-neighborhood of $x_i$. If the number of points inside this radius meets or exceeds a predetermined minimum threshold value, the point is deemed a core point, and all neighboring points within the radius are incorporated into its cluster. The clusters are expanded recursively by incorporating neighboring core points and their related border points. This procedure continues until all points are either assigned to a cluster or designated as noise points if they don't belong to any cluster.

In this study, each data point $x_i$ is a three-dimension vector indicating a census tract's exposures to three kinds of hazards, i.e., $x_i = (MEI_{ap}^i, MEI_t^i, MEI_h^i)$, and the distance metric $d(.,.)$ is defined as the Euclidean distance between data points.

## 2.4 Hazard exposure disparity for socially vulnerable populations

We conducted the T-test to examine the presence of disparity in the extent of mobility-based hazard exposure for socially vulnerable populations. T-test is a statistical hypothesis to determine whether there is significant difference between the means of two groups of data. It is calculated by taking the difference between the means of the two groups and dividing it by the standard error of the difference between the means. In this study, the proportions of vulnerable subgroups (i.e., minority, below poverty, Section 2.1) among different hazard exposure regions are quantified, and the T-test is then applied for statistical comparisons.

## 3. Results

### 3.1 Patterns of hazard exposures

In this study, the regions with high hazard exposure are also called direct exposure regions, which denotes the population living within the regions experience direct high hazard exposure due to their residence; In contrast, other regions with mobility-based exposure index (MEI) $> 0$ are defined as latent exposure regions, indicating that the residents in these regions, despite living in direct exposure regions, still bear latent hazard exposure due to their visits to places in other high-hazard regions.

The distributions of MEIs show great unevenness across the studied region (Figure 2 (b)-(d)). For all studied hazards (air pollution, toxic exposure, heat), the MEIs show patterns of "bimodal distribution, with MEI clustering around separate values and peaking at the two ends. Specifically, the direct exposure regions have higher exposure than the latent exposure regions. The mean values of $MEI_{ap}$, $MEI_t$, and $MEI_h$ for the direct exposure regions are 98.9%, 95.3%, and 96.5%, as opposed to 4.4%, 7.8%, and 2.5% for the latent exposure regions, confirming the residential location as the major determinant for hazard exposures. The bi-modal distribution MEI reveals the presence of significant gap in the extent of environmental hazard exposure. One group of individuals spends more than 80% of their time in places with high hazard areas and another group spends less than 20% of their time in high-hazard areas. The presence of this significant divide in MEI is an indicator of environmental injustice and motivates additional analysis to examine disparities for socially vulnerable populations.

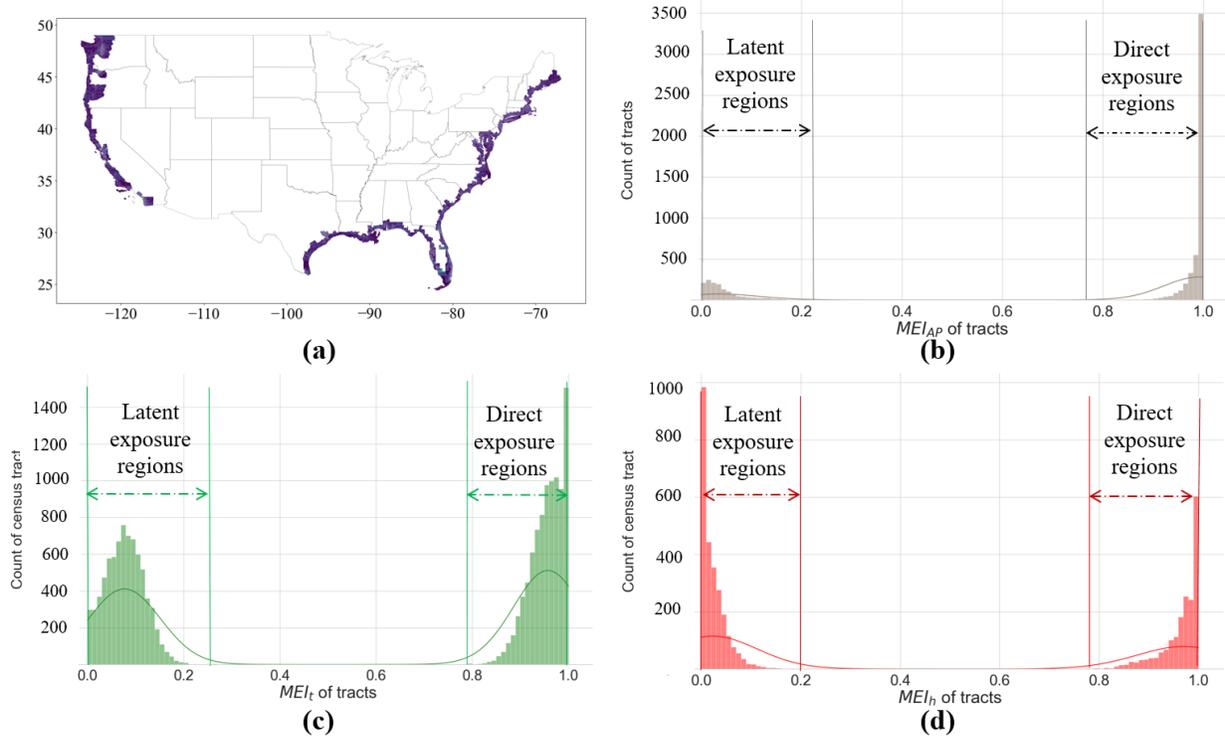

**Figure 2**. Distributions of MEI for different hazard types. (a) The geographical distributions of the study regions, along the coastal areas in U.S. (b)-(d) Histogram representation of the $MEI_{ap}$, $MEI_h$, $MEI_t$ across the study regions. For all the three studied hazards, the $MEIs$ follow bimodal distribution with peaks at both ends, demonstrating distinct hazard exposures between direct and latent exposure regions, and that residential locations are the major determinant for people's hazard exposures. Note that tracts with no hazard exposure are excluded from the plot.

To decipher the influence of mobility on hazard exposures, we further identify the hazard exposures induced by the residents' visits to non-home regions (the regions other than their home census tracts), for both direct and latent exposure regions (Figure 3). For the direct exposure regions, the mean values of traveling-induced $MEI_{ap}$, $MEI_t$, and $MEI_h$ are respectively 15.3%, 12.2%, and 12.0% greater than those of the latent exposure regions, with increased MEI values of 3.2%, 7.8%, and 1.7%, respectively. These findings indicate that: (1) individuals residing in regions with direct hazard exposure not only experience a greater degree of hazard exposure owing to their residence in a high-hazard census tract, but they also face an higher risk of exposure to hazards during visits to places in nearby high-hazard regions than individuals residing in areas with mere latent exposure; (2) disregarding the impact of mobility on environmental hazard exposure may result in the under-estimation of hazard exposure risks, which is particularly evident in regions with direct exposure, where, on average, more than 10% of hazard exposures may be underestimated. The fact that people residing in high-hazard areas have a greater exposure due to their mobility can be explained based on spatial clustering of hazards and laws of human mobility. According to the laws of human mobility, the frequency of visits to places has an inverse relationship with distance to home (Schläpfer, Dong et al. 2021). Since environmental hazards are spatially clusters, a high-hazard census tract is likely to be surrounded by other high-hazard census tracts as well. Hence, when people visit places in non-home census tracts, it is more likely to visit places in neighboring census tracts, which have similar levels of hazard exposure, more frequently.

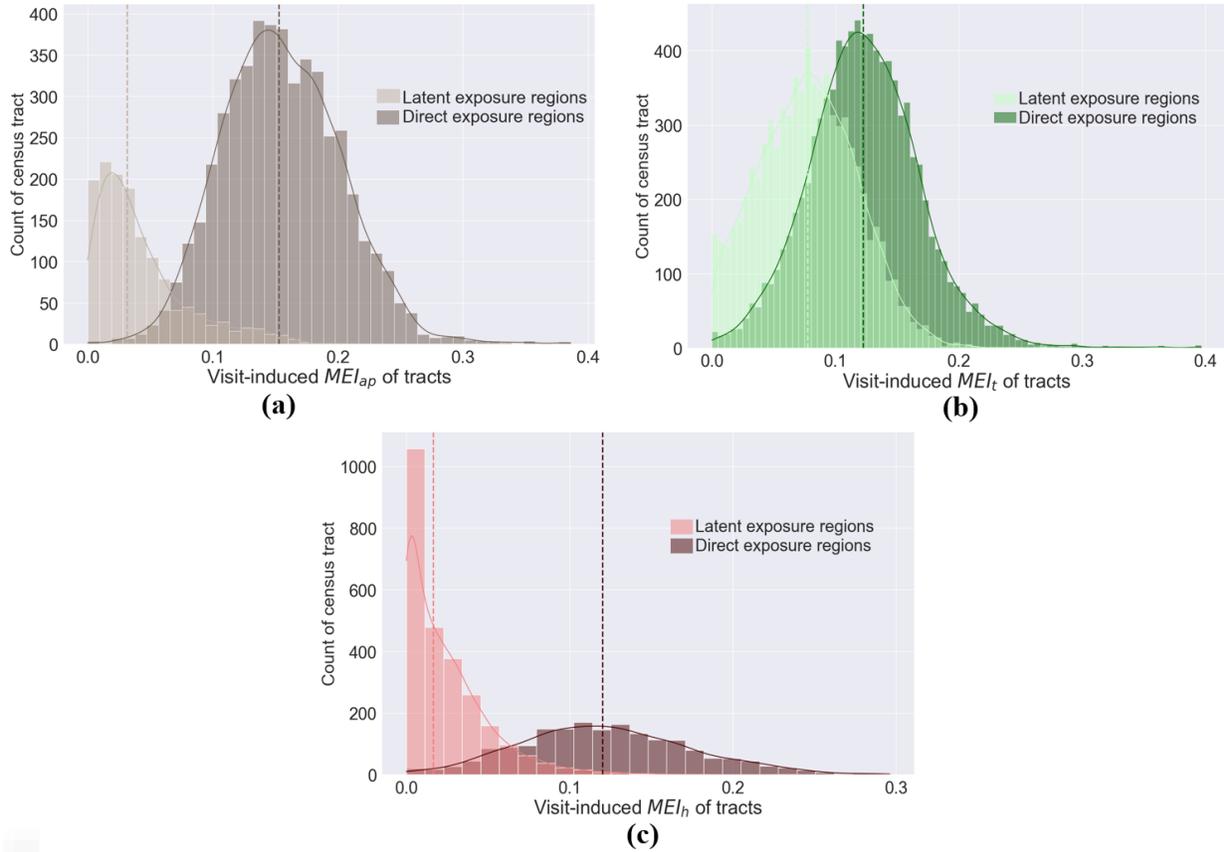

**Figure 3**. Comparisons of MEI due to populations' visits to non-home regions (that is, other than their home census tract), between direct and latent exposure regions. Individuals residing in regions with direct hazard exposure experience a greater exposure to hazards due to visits to nearby regions than individuals residing in areas with mere latent exposure.

In addition, variations in MEIs are evident across diverse geographical areas. As shown in Figure 4 (a)-(b), eight distinct clusters/categories have been identified according to the hazard exposures experienced by each tract (refer to Section 2.3). Every category exhibits a unique degree of exposure to different hazard types. A significant proportion of tracts (73.3%) face the risk of multiple (two or three) high-hazard exposures. Notably, the category characterized by high exposure to all three types of hazards encompasses the largest number of tracts within the study areas, constituting 32.7% of the tracts in total.

Also, the co-occurrence of multiple types of hazards in specific areas suggests potential correlations between these hazards. Further analysis of the mobility-based hazard exposure index (Figure 4 (c)) reveals significant positive correlations at a significance level of 0.01 between each pair of $MEI_{ap}$, $MEI_t$, and $MEI_h$. This result reveals another important aspect of environmental injustice. People in high hazard areas typically have high exposure to two or more hazard types.

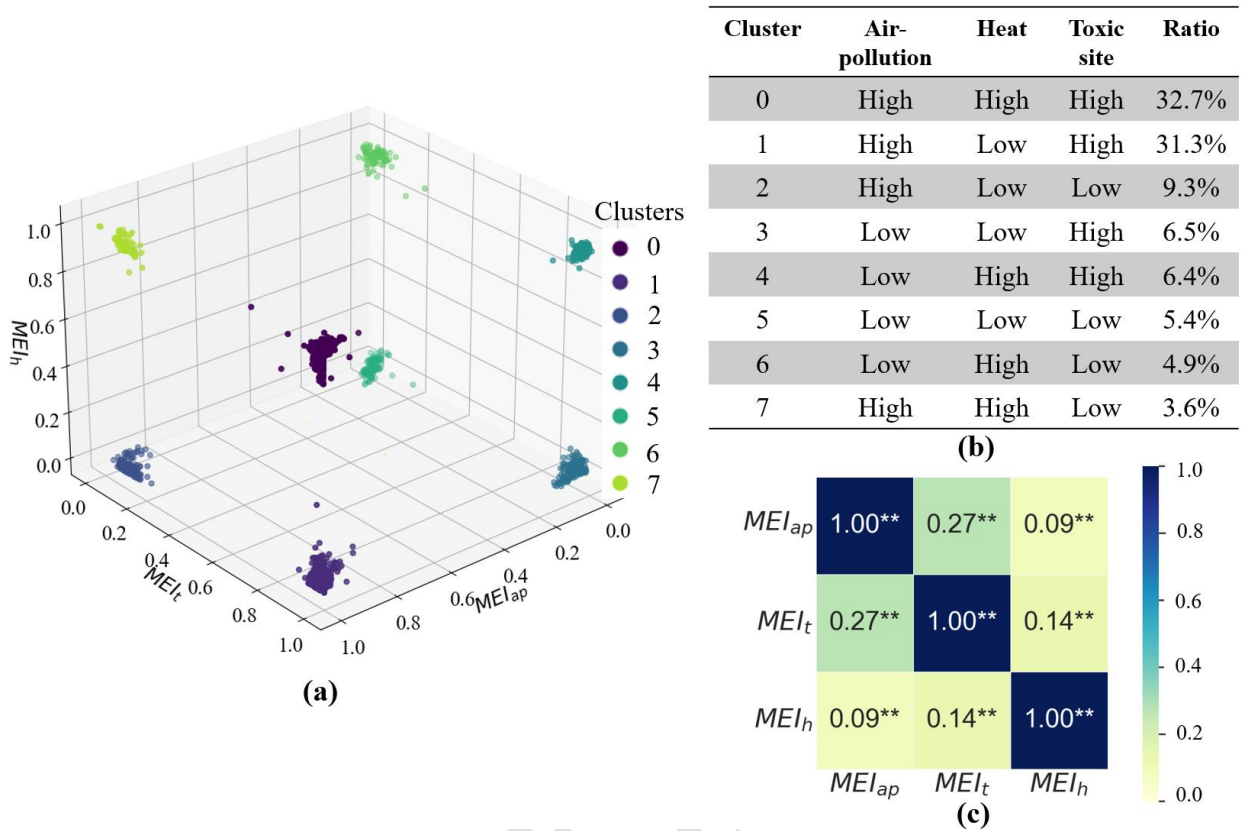

**Figure 4**. Classification of census tracts based on MEI patterns for all hazard types, using DBSCAN. (a) Representation of the census tracts in three-dimensional space. Each point represents a census tract. The coordinates of the point are the tract's possessed value of $MEI_{ap}$, $MEI_t$, and $MEI_h$. The color indicates the tract's category. (b) The count of tracts and occupation in respective hazards for each category. (c) The analysis of correlations between hazard pairs.

### 3.2 Latent exposure areas

Individuals residing in regions without immediate hazard exposure may still encounter persistent and chronic hazard exposure due to their outbound visits to other high-hazard regions. Figure 5 plots affected populations that have certain levels of hazard exposure in the latent exposure areas. The results reveal that, among the study areas (239 coastal counties in the United States), 1.16 million people live in regions without direct exposure to toxic sites but still spend more than 10% of their life activities in areas in proximity of toxic sites due to visits to other areas. In comparison, 0.13 million and 0.66 million individuals encounter similar degrees of latent exposure to air pollution and heat, respectively. When the threshold is set at 5%, populations with latent exposure to air pollution, heat, and toxic sites increases to 1.99 million, 1.26 million, and 27.11 million, respectively. While spending 5% of life activity times in high hazard areas may not seem significant, persistent exposure for this duration could have dire health and wellbeing impacts. This result also highlights that exposures to toxic sites constitute the largest source of latent risk among the three types of investigated hazards.

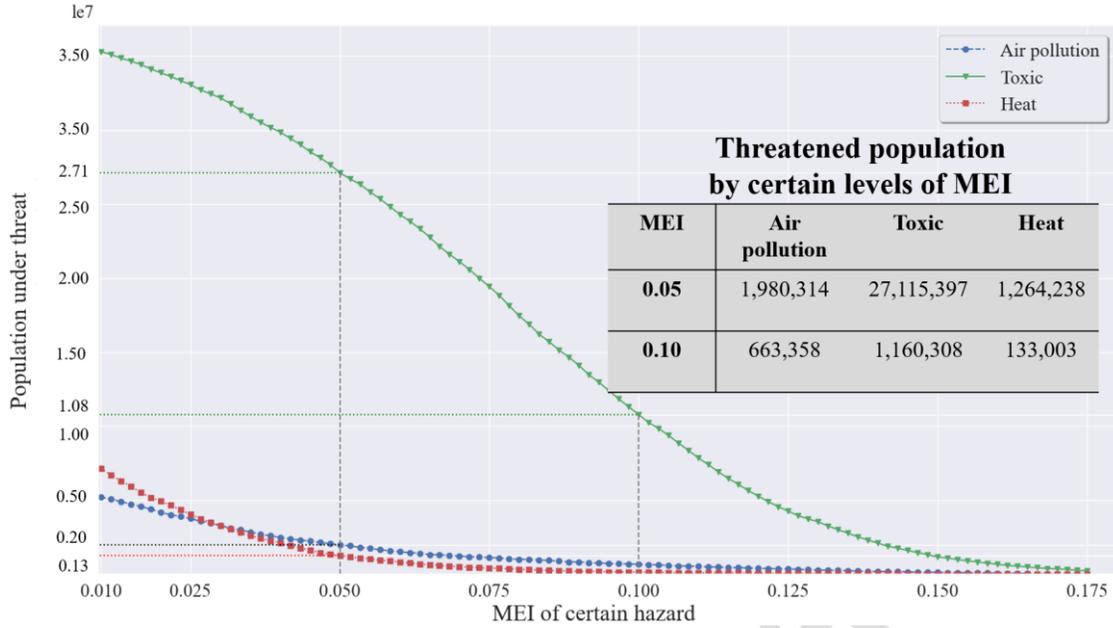

**Figure 5**. Populations affected by different hazard types in the latent exposure areas. The x-axis is the MEI, representing the proportions of total time in high hazard areas. The y-axis is the number of the affected population based on population census data at census-tract level.

Some areas experience the combined effects of multiple latent hazard exposures. For example, among the 239 counties examined in our study, Queens County in New York City was identified to have census tracts with latent exposure to all three investigated hazards. Figure 6 depicts that over 54, 176 and 118 tracts in Queens County have an $MEI_{ap}$, $MEI_t$, and $MEI_h$ exceeding 5%, which implies that residents in these areas are exposed to air pollution, toxic and heat hazard exposures for more than 10% of their time due to travel to other high-hazard areas. Specifically, there are 19 tracts with latent exposure to all three hazard types, subjecting more than 60,000 persons to the compounded effects of these latent hazard exposures.

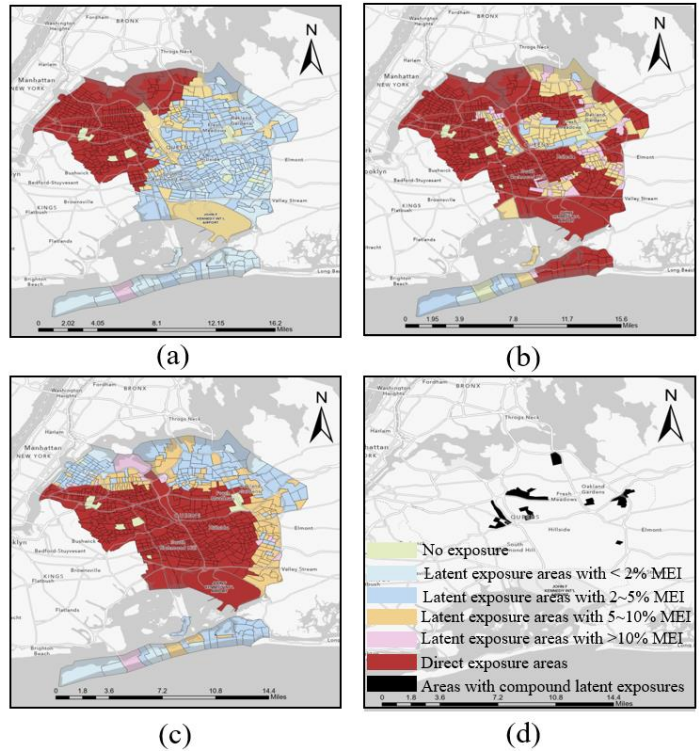

**Figure 6**. The various hazard exposures in Queens County in New York City, (a) air pollution, (b) toxic site exposure, (c) heat, (d) compound (the areas affected by all the three hazards)

## 3.3 Disparity in hazard exposures

The results show significant disparity in MEI across different sub-populations, raising new concerns for environmental injustice. Figure 7 shows that, for all the three hazard types, ethnic minority sub-populations and sub-populations with a greater degree of poverty have a larger MEI values. A closer demographic comparison between the high-hazard exposure areas and other areas is shown in Table 1. The results show that, compared with the average level, both the direct and latent exposure areas generally have higher concentrations of socially vulnerable community (i.e., minority and below-poverty population), with one exception for the latent air-pollution exposure areas. The results provide data-driven evidence that below poverty and ethnic minority residents have a greater mobility-based environmental hazard exposure (both in direct exposure areas and latent exposure areas).

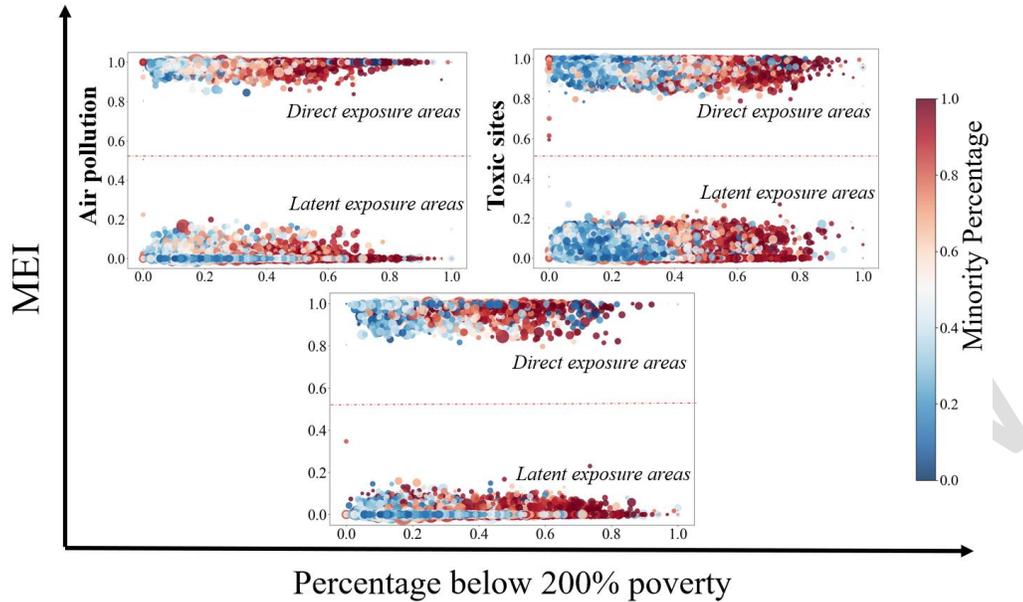

**Figure 7**. The hazard exposure (measured by MEI) versus community vulnerability indicator at the census tract-level. Each data point in the chart represents a tract. The x-axis represents the tract's percentage of population below 200% poverty; the y-axis represents the tract's the MEI for various hazards; the color of each point is rendered by the tract's percentage of minority and the size of each point is scaled proportional to the tract's population.

**Table 1**. The statistics of vulnerability representativeness across different areas. The numbers larger than the average of all tracts are colored in red texts and ** represents T-test significant at 0.01 level.

| | Average of all tracts | Direct exposure areas | | | | Latent exposure areas | | | |
|---|---|---|---|---|---|---|---|---|---|
| | | Air pollution | Toxic | Heat | Compound | Air pollution | Toxic | Heat | Compound |
| **Percentage below 200% poverty** | 29.0% | 30.4%** | 31.8%** | 32.7%** | 35.9%** | 26.0%** | 33.8%** | 28.5%** | 53.4%** |
| **Percentage of minority** | 49.8% | 58.0%** | 55.1%** | 58.4%** | 64.8%** | 48.9%** | 59.3%** | 53.9%** | 81.0%** |

Another noteworthy finding is that areas exposed to all three hazards exhibit particularly high concentrations of vulnerable communities, with both below-poverty and minority populations being disproportionately represented. In areas with compound direct exposures, the below-poverty and minority populations account for 35.9% and 64.8% of the total population, respectively. For areas with compound latent exposures, these proportions rise even higher, to 53.4% and 81.0%. This indicates that vulnerable communities, especially those experiencing poverty and with minority populations, are disproportionately burdened by the compounding effects of multiple hazards. These findings highlight the urgent need for targeted interventions, policies, and resources that specifically address the unique challenges faced by these communities in order to reduce their vulnerability and promote environmental justice and equitable living conditions, which will be further discussed.

## 4. Discussion

### 4.1 Spatial autocorrelation and human mobility potentially create traps in direct exposure areas

Our results reveal a more nuanced understanding of the environmental risks endured by individuals living in areas with direct hazard exposure, showing that their exposure level extends beyond what was previously know and the risk from a seemingly insignificant exposure level could have dire health consequences. These individuals, while already susceptible due to residence in close proximity to hazard zones, face an additional risk as they visit places within nearby areas.

The observed increase in hazard exposure due to visits to places in nearby areas could be attributed to the spatial autocorrelation of high-hazard areas. Essentially, regions that are geographically close are more likely to exhibit similar hazard characteristics due to shared environmental and development and land use factors. For instance, air pollution levels might be high across a cluster of census tracts due to the prevalence of industries or dense traffic routes, not just in the immediate vicinity of an individual's residence. Similarly, a region with high heat exposure due to urban heat-island effect might be surrounded by areas with similar characteristics due to shared climatic conditions and urbanization patterns. This spatial autocorrelation means that individuals residing in regions with direct hazard exposure are likely to encounter similar, if not identical, hazards in their proximal surroundings when they travel.

Coupling with the spatial autocorrelation of hazards, human mobility law related to decay effect (i.e., "individuals are more likely to visit nearby regions") is possibly another cause for accentuated traveling-induced hazard exposures (Chen, Shi et al. 2019, Liu, Zhang et al. 2021). The human mobility patterns are greatly influenced by the urban environment and transportation (Yin, Li et al. 2018, Shen, Shi et al. 2022, Wang, Yang et al. 2022, Yin, Wu et al. 2023). Given that people tend to frequent places that are closer to their residence, their mobility patterns essentially increase their dwell time with high-hazard environments if they live in high-risk areas. If their home and nearby regions share similar hazard characteristics due to spatial autocorrelation, their likelihood of exposure increases with each trip they make within this radius. This highlights the importance of taking into account both the geographical distribution of hazards and human mobility patterns when assessing the true extent of hazard exposure risk. In other words, the combined effects of spatial autocorrelation of high-hazard areas and human mobility patterns created environmental hazard traps for residents living in these areas.

### 4.2 Examining the intersection of hazards interplay and environmental injustice

The findings from this study underline two significant aspects of environmental hazard exposure: the interrelated nature of different hazards and the disproportionate impact on vulnerable communities. First, the observed co-occurrence and significant positive correlations between the MEIs for the three hazard types suggest that these environmental threats interact and potentially exacerbate each other, creating a complex risk landscape. This interplay between air pollution, toxicity, and heat hazards could contribute to the heightened risk for individuals living in and visiting these high hazard areas. Understanding this interrelation is critical, as it highlights the need for integrated environmental hazard mitigation strategies for tackling hazards, rather than addressing them in isolation.

Second, the findings underscore a pressing environmental justice issue. First, the bi-modal distribution of MEI values suggests a large divide in the extent of exposure of individuals for all three hazards types. This large divide provides data-driven evidence regarding magnitude of environmental injustice in these counties. Second, it is alarming that below-poverty and minority populations are disproportionately represented in areas exposed to all three hazards. The higher representation of these sub-populations in areas with compound exposures is indicative of significant disparity in the distribution of environmental risks. These communities are shouldering an unjust burden, being more affected not only by individual hazards but also by the compound effects of all three hazards through both direct and latent exposures. This level of exposure exacerbates their existing vulnerability and poses significant challenges to their health and well-being. In light of these findings, it is crucial to emphasize the need for targeted interventions and policies that specifically address the unique challenges faced by these communities. Efforts should be made to reduce their vulnerability, promote environmental justice, and achieve equitable living conditions. This could involve, for instance, enhancing infrastructure in these areas to reduce hazard exposure, implementing stricter regulations on pollution sources, or providing better access to healthcare and resources for coping with hazards (Liu and Mostafavi 2023, Rajput, Jaing et al. 2023). Furthermore, these results underline the importance of incorporating an environmental justice lens in hazard management and urban planning policies to ensure a fair distribution of environmental risks

### 4.3 Latent exposure: a chronic threat

The findings showed a significant number of residents in the study regions spend 5%-10% of their total life activities in high hazard areas while their residence is not in high-hazard areas (i.e., residents of latent exposure areas). While these values are significantly less than the MEI values of direct exposure areas, persistent exposure to these hazards for 5%-10% of total weekly activity times (i.e., MEI > 5%) could compound over time to a significant amount of exposure over years and cause dire health impacts. While the environmental hazard exposure of residents in direct exposure areas could cause acute health effects, the effects on residents in latent hazard areas is chronic and compound over time. This finding also implies the spillover effects of environmental hazard mitigation strategies that would extend the benefits of exposure reduction and hazard mitigation policies and actions to residents residing outside the high-hazard areas by reducing their latent exposure caused by human mobility.

## 5. Concluding Remarks

Environmental hazards pose significant threats to human health. Traditional approaches for hazard exposure evaluation have mostly focused on individuals' home residence location and neglect hazard exposure due to people's mobility and dwell time in other regions. In this study, we create a mobility-based approach for quantifying the extent of environmental hazard exposure by utilizing fine-grained large-scale human mobility dataset.

The proposed mobility-based hazard exposure index and the findings from the analysis of 239 U.S. coastal counties have multiple important contributions. First, by considering the extent of hazard exposures related to population activities and visits to places in other regions, this study provides a more reliable measure for quantifying environmental hazard exposures. The findings show that overlooking the influence of human movement on exposure to environmental hazards can potentially lead to an underestimation of these risks. This issue is especially apparent in areas that face direct exposure, where, on average, there may be an underestimation of over 10% of hazard exposure instances. Second, the results showed that the combination of spatial clustering of high-hazard areas and distance-decay law of human mobility has led to creation of environmental hazard traps in which residents residing in high-hazard areas bear addition 10% exposure to environmental hazards due to their life activities and visitation to places in other areas.

Third, the findings reveal a significant divide in the extent of environmental hazard exposure in residents of communities. The bi-modal distribution of MEI values for both direct exposure and latent exposure areas provide data-driven evidence for the severity of environmental injustice issues. In addition, the findings showed poor and ethnic minority residents are disproportionately exposed to all the three environmental hazard types examined in this study. Also, these vulnerable populations are exposed to more than two environmental hazard types in most census tracts, which compounds the adverse health impacts of these hazards. These findings provide a deeper understanding of the extent of environmental injustice in U.S. communities. Finally, the study enables the examination of latent exposure to environmental hazards caused by visitation to places outside the home census tracts. The results show that millions of U.S. residents in the counties studied spend 5% to10% of their weekly life activities dwelling in places with high hazard exposure. This latent exposure could compound over time and cause a chronic threat to the health of populations who are not living in high-hazard areas. A substantial portion of the population might be inadvertently exposed to various hazards due to their daily routines, despite residing in areas perceived as safe. These findings offer important data-driven insights to public health officials, urban planners, and environmental policy makers regarding the extent of environmental hazard exposure in different areas of a community in formulating targeted policies to reduce the dire environmental injustice shown in this paper.

The results if this research suggest several worthwhile directions to explore in the future. First, the proposed approach in this work could be expanded to include other kinds of environmental hazards. By exploring more diverse hazards, researchers could provide a deeper understanding of the different ways in which mobility intersects with various forms of hazard exposure. Second, there lies an intriguing opportunity to model the relationship between mobility-based hazard exposure and certain health outcomes, such as disease incidence, life expectancy, and mortality rates. A data-driven exploration of these relationships could yield invaluable insights into the public health implications of environmental hazards in conjunction with population activities and human mobility. Also, future studies could adopt the mobility-based approach proposed in this study to examine the effects of future urban development and climate change on population exposure to environmental hazards. For example, expansion of cities changes population density, modifies spatial distribution of hazards and alters human mobility. The analysis of combined effects of city development and expansion using the mobility-based approach proposed in this study could inform integrated urban design strategies to address environmental hazards and public health impacts as cities continue to grow and expand.

**Acknowledgement**

The authors would like to acknowledge funding support from the National Science Foundation CRISP 2.0 Type 2 No. 1832662: Anatomy of Coupled Human-Infrastructure Systems Resilience to Urban Flooding: Integrated Assessment of Social, Institutional, and Physical Networks. Any opinions, findings, and conclusions or recommendations expressed in this research are those of the authors and do not necessarily reflect the view of the funding agencies.

**Data availability**

The human mobility datasets are provided by Spectus Inc. The other datasets that support the findings of this study are available from upon reasonable request.